\documentstyle[aps,twocolumn,epsfig]{revtex}

\begin{document}

\narrowtext

\noindent 

\noindent {\bf Tuning the trip to KPZ asymptopia}
\vspace{4mm}

The paradigmatic stochastic partial differential equation proposed by
Kardar, Parisi and Zhang (KPZ) 
has, in the past decade, become the cornerstone in our understanding of
kinetic roughening phenomena; mathematical links to directed polymers in
random media (DPRM), a baby version version of the spin-glass problem
exactly soluble in $d=1,$  as well as driven diffusive systems, doing
much to intensify community interest in the strong-coupling,
nonperturbative scaling behavior of this fundamental equation~\cite{HHZ}.
Central to the KPZ dogma is the implicit assumption of universality;
i.e.,  the equation provides a bare-bones, but catch-all  continuum
description that transcends all irrelevant microscopic details,
particularly those associated with the lattice.    Indeed, within the
immediate context of nonequilibrium stochastic growth, the KPZ equation
reigns supreme- its very broad universality has been
numerically established well beyond characteristic scaling exponents, to
include finite-size corrections,  amplitude ratios and full probability
distributions~\cite{krug}. 
This Comment arises from a
recent dialog concerning the suggestion of possible KPZ nonuniversality, initiated
by Newman and Swift~\cite{swift}, who numerically integrate the KPZ equation in
the strong-coupling limit, where their prescription reduces to a
zero-temperature DPRM analysis.  For gaussian noise distributions,
they readily retrieve accepted values for the KPZ energy
fluctuation exponent $\beta$ in $d=1-4$.   For bounded, symmetric, and innocuous
distributions of the form: $P(\varepsilon)\sim(1-|\varepsilon|)^a,$ however, they notice 
increasingly
greater deviations as $a\rightarrow-1.$  In fact, the discrepancies grow more
severe in higher dimensionalities.  This behavior is easily seen as a
consequence of the close blood relation between the DPRM and its
near cousin, directed percolation (DP), as emphasized by several authors
~\cite{HH,Tang}, who point out the important role of falling percolation thresholds with increasing 
dimensionality.  Our purpose here is clarify the essential features of this tale, insuring that 
crucial bits of evidence remain atop the table.  In this manner, one might hope to discover 
the optimal road to KPZ asymptopia.   
  
        Much can be gleaned, already, using truncated gaussian noise,
restricting ourselves to either positive or negative bond energies,
thereby breaking the symmetry intrinsic to the full distribution.
Interestingly,  these demi-gaussians yield strikingly different scaling
behaviors in the zero-temperature DPRM global minimization procedure.
For a  positive-bond demi-gaussian, $P(\varepsilon > 0)\sim e^{-\varepsilon^2/2}$, 
we obtain an anomalous 
exponent that
is much too small, though entirely reminiscent of Newman-Swift
results; e.g., in a modest, but highly representative, simulation of the 2+1 dimensional DPRM, 
involving $10^5$ realizations of the random energy landscape and directed paths of length $10^2$ 
steps, the extracted fluctuation exponent, $\beta\approx0.116,$ is well below the accepted value
$0.24^+$~\cite{LH}.  Naively, this result appears counterintuitive, since a positive demi-gaussian 
 is strongly biased low, possessing a relatively high percentage of favorable, low energy bonds attractive
to the directed polymer.  Yet, it is precisely this abundance of riches which is responsible
for a near ground-state degeneracy that greatly delays the DPRM's true approach to asymptopia. 
By contrast, for a negative-bond demi-gaussian,
$P(\varepsilon <0)\sim e^{-\varepsilon^2/2}$, which is biased high, we retrieve 
excellent KPZ scaling; i.e.,  a simulation, 
otherwise identical to that just mentioned, now produces an exponent $\approx0.245$, 
even for these 
rather short optimal paths!  The relative scarcity of particularly favorable bonds permits, 
somewhat 
unexpectedly, an accelerated convergence to asymptopia.  Surprisingly, a simulation employing 
the full 
gaussian {\it actually fares less well,} producing an exponent $\approx 0.216$ under the same conditions.   

Splitting the
Newman-Swift bimodal ($a=-1$) distribution into two pieces unearths, of course, the very same
pathology- an immediate DPRM vestigial link to directed percolation ~\cite{HH,Tang}.   Biasing high, 
with positive bond energies only,
generates excellent KPZ scaling; as impressive, in fact, as the previous demi-gaussian!  By 
contrast, biasing low yields a disasterously small effective exponent.  The dangers inherent in the 
Newman-Swift class of distributions, although most apparent for $a<0$, are also clearly present for 
$a\ge 0,$ the essential difference being one of degree, rather than kind, since all such bare 
probability distributions lie within the DPRM basin of attraction, albeit at different distances 
from the inevitable fixed point function. Indeed, the kernel of the Newman-Swift numerical mystery 
well predates their paper, and can be found originally in the work of Kim, Bray and Moore 
~\cite{kbm}, wherein it is revealed that a flat distribution ($a=0$), while the gaussian's peer 
within the 1+1 DPRM context, seriously underperforms the latter in 2+1 dimensions, for which the DP 
threshold has fallen below one-half~\cite{grass}.  Furthermore, earlier work of Derrida and Griffiths~\cite{dg} 
had indicated that a sufficiently asymmetric, discrete bare probability distribution would land the 
model well outside the KPZ universality class, controlled by the DP fixed point.  Problems 
inevitably arise whenever there is too much integrated weight in the low end of the chosen DPRM bond 
distribution~\cite{HH}. Such is the price paid for a surprising, but deep blood connection to 
directed percolation.

\vspace{3mm}
\noindent This research was supported via NSF DMR-9528071.
\vspace{5mm}

\noindent T. Halpin-Healy and Rocky Novoseller

\noindent Physics Department, Barnard College

\noindent New York, New York 10027-6598

\end{document}